\documentstyle[12pt]{article}

\renewcommand{\thefootnote}{\alph{footnote}}

\newcommand{\beqn}{\begin{equation}}
\newcommand{\eeqn}{\end{equation}}
\newcommand{\beqarr}{\begin{eqnarray}}
\newcommand{\eeqarr}{\end{eqnarray}}

\newcommand{\sj}{\sigma_j}
\newcommand{\di}{\partial_i}
\newcommand{\dj}{\partial_j}
\newcommand{\Ut}{\widetilde{U}}

\begin{document}

\begin{titlepage}

July 1998         \hfill
\begin{center}
\hfill    UCB-PTH-98/38 \\
\hfill    LBNL-42104	 \\

\vskip .15in
\renewcommand{\thefootnote}{\fnsymbol{footnote}}
{\large \bf Super Yang-Mills on the Noncomutative 
Torus}\footnote{To appear in the Arnowitt Festschrift Volume
``Relativity, Particle Physics, and Cosmology'', 
Texas A\&M University, April 1998,
published by World Scientific}\footnote{This work was supported in
part by the Director, Office of Energy Research, Office of High Energy and 
Nuclear Physics, Division of High Energy Physics of the U.S. Department of 
Energy under Contract DE-AC03-76SF00098 and in part by the National Science
Foundation under grant PHY-95-14797}
\vskip .15in

Bogdan Morariu\footnote{email address: bmorariu@lbl.gov} and 
Bruno   Zumino\footnote{email address: zumino@thsrv.lbl.gov}
\vskip .15in

{\em 	Department of Physics			\\
	University of California				\\
				and									\\
	Theoretical Physics Group			\\
	Lawrence Berkeley National Laboratory	\\
	University of California				\\
	Berkeley, California 94720}
\end{center}
\vskip .15in

\begin{abstract}
After a brief review of matrix theory compactification 
leading to noncommutative supersymmetric Yang-Mills gauge theory, 
we present solutions for the fundamental and adjoint
sections on a two-dimensional twisted quantum torus in two different gauges. 
We also give explicit transformations connecting different 
representations which have appeared in the literature. Finally we 
discuss the more mathematical concept of Morita equivalence 
of $C^{*}$-algebras as it applies to our specific case.
\end{abstract}
\end{titlepage}
\renewcommand{\thepage}{\roman{page}}
\setcounter{page}{2}
\mbox{ }

\vskip 1in

\begin{center}
{\bf Disclaimer}
\end{center}

\vskip .2in

\begin{scriptsize}
\begin{quotation}
This document was prepared as an account of work sponsored by the United
States Government. While this document is believed to contain correct
 information, neither the United States Government nor any agency
thereof, nor The Regents of the University of California, nor any of their
employees, makes any warranty, express or implied, or assumes any legal
liability or responsibility for the accuracy, completeness, or usefulness
of any information, apparatus, product, or process disclosed, or represents
that its use would not infringe privately owned rights.  Reference herein
to any specific commercial products process, or service by its trade name,
trademark, manufacturer, or otherwise, does not necessarily constitute or
imply its endorsement, recommendation, or favoring by the United States
Government or any agency thereof, or The Regents of the University of
California.  The views and opinions of authors expressed herein do not
necessarily state or reflect those of the United States Government or any
agency thereof, or The Regents of the University of California.
\end{quotation}
\end{scriptsize}

\vskip 2in

\begin{center}
\begin{small}
{\it Lawrence Berkeley National Laboratory is an equal opportunity employer.}
\end{small}
\end{center}

\newpage
\renewcommand{\thepage}{\arabic{page}}

\setcounter{page}{1}
\setcounter{footnote}{0}


\section{Introduction}
It was conjectured in \cite{BFSS} that the infinite momentum frame 
description of M-theory is given by the large $n$ limit of 
supersymmetric quantum mechanics (SQM) \cite{CH,FL,BRR},
obtained as the dimensional reduction of the $9+1$ dimensional 
$U(n)$ supersymmetric Yang-Mills (SYM) gauge field theory.
Shortly afterwards Susskind  took this a step further~\cite{LS}, 
conjecturing that
the discrete light cone quantization (DLCQ) of M-theory is
equivalent to the finite $n$ matrix theory.

Toroidal compactification of M-theory can then be obtained by first
considering matrix theory on the covering space and then imposing 
a periodicity condition on the matrix variable \cite{BFSS,WTc,GRT}, also 
known as the quotient condition. The result is a SYM field theory on a 
dual torus.

If we consider the DLCQ of M-theory and compactify on a torus $T^d$ for
$d \geq 2$ there are additional moduli coming from the three-form
of 11-dimensional supergravity. For example, if we compactify on $T^2$
along $X_1$ and $X_2$ then $C_{-12}$ cannot be gauged away, and is a modulus
of the compactification. It was conjectured in \cite{CDS} that
turning on this modulus corresponds to deforming the SYM theory
on the dual torus to a noncommutative SYM on a quantum torus~\cite{CR} with
deformation parameter $\theta$ given by
\[
\theta=\int~C_{-12}\,dX_{-} dX_1 dX_2.
\]
Evidence for this conjecture comes from  comparison of the BPS mass spectra 
of the two theories and of their duality groups. Further evidence and 
discussions of this conjecture followed 
in~\cite{DH,PWW,PW,C,CK,KO,AAS,PMH,AS1,AS2}.

In Section~\ref{SYM} we present a review of matrix theory compactification
leading to noncommutative SYM gauge theory on trivial quantum bundles. 
We follow the 
elementary treatment of~\cite{PMH} with an emphasis on giving 
explicit formulae that closely resemble the commutative case. 
We further
present an explicit realization of the
algebra of the quantum torus ${\cal A}(T^2_{\theta})$ in terms of 
quantum plane coordinates.

In Section~\ref{TQB} we introduce non-trivial quantum bundles
as in~\cite{PMH} corresponding to compactified 
DLCQ of M-theory in the presence
of transversely wrapped membranes.
We also explain 
in some detail how to solve the boundary conditions for sections in the 
fundamental and adjoint quantum bundle. 
Finally using the special form of the transition 
functions in the given gauge we find an equivalent but simpler form of
the general solution for fundamental sections.

In Section~\ref{modules}  we  discuss the more abstract 
language of projective 
modules, as presented in~\cite{CDS} and references therein, and we then 
give the 
explicit map between this formulation and the more elementary 
formulation in \cite{PMH}. 
We also explain the notion of Morita equivalence~\cite{R1,CDS,AS1,AS2,LLS}
applied to our specific case. For an expanded coverage of noncommutative
geometry see~\cite{CON} and for a brief description see~\cite{DB}.

Finally in Section~\ref{gauge-trans} we discuss the
general theory of gauge transformations
on the noncommutative torus and find an explicit gauge transformation that
trivializes one of the transition functions. With trivial transition
functions T-duality transformations take the standard form, allowing us to
interpret the gauge field as D-strings on the dual torus.

\section{Review of Matrix Compactification}
\label{SYM}

In this section we present a review of matrix theory compactification 
closely following the description given by Ho in~\cite{PMH}. The $P_{-}=n/R$
sector of the DLCQ of uncompactified M-theory is given by the $U(n)$ 
SQM~\cite{CH,FL,BRR}
whose action in the temporal gauge is given by
\beqn
S=\frac{1}{2R} \int dt\, {\rm Tr}\left(\dot{X}^{\mu}\dot{X}_{\mu} +
\sum_{\mu > \nu}[X_{\mu},X_{\nu}]^2+
i \Theta^{T} \dot{\Theta} - \Theta^{T} \Gamma_{\mu}[X^{\mu},\Theta]
 \right)    \label{action}
\eeqn
where $\mu,\nu=0,\ldots,9$.
We will compactify matrix theory 
on a rectangular 2-torus of radii
$R_1$ and $R_2$.  First let us
consider  matrix variables on the covering space 
and impose the quotient condition
\beqarr
U_i^{-1} X_j U_i &=& X_j + 2 \pi R_j \delta_{ij},  \label{quotient} \\ 
U_i^{-1} X_a U_i &=& X_a,    \nonumber   \\
U_i^{-1} \Theta ~U_i &=& \Theta,~~~i,j=1,2~~~a=3, \ldots,9.     \nonumber
\eeqarr
Here the $U_i$\/ are unitary operators. The consistency of these equation 
requires 
\[
U_1 U_2 = e^{2\pi i \theta} U_2 U_1.
\]
Before solving the quotient condition (\ref{quotient}), it is convenient 
to introduce two more unitary operators $\Ut_i,~~ i=1,2$ which commute 
with the $U_i$'s and satisfy the relation
\beqn
\Ut_1 \Ut_2 = e^{-2\pi i \theta} \Ut_2 \Ut_1. \label{Ut1Ut2}
\eeqn
One way to realize this algebra is by using canonical variables $\sigma_i$
satisfying
\beqn
\left[  \sigma_1, \sigma_2 \right] = 2 \pi i \theta. \label{ss}
\eeqn
Then $\Ut_i \stackrel{\rm def}{=} e^{i \sigma_i}$\/ 
satisfy (\ref{Ut1Ut2}). The variables 
$\sigma_i$ are noncommutative coordinates on the quantum plane which is the
covering space of the quantum 
torus. The algebra of functions on the quantum torus denoted
${\cal A}(T^2_{- \theta})$ is generated by $\Ut_i$. Similarly
the $U_i$ operators generate the algebra denoted ${\cal A}(T^2_{\theta})$.
To realize them we introduce partial derivative operators 
on the quantum plane, satisfying the following algebra
\beqn
\left[  \di, \sj \right] = \delta_{ij}, \label{ds}
\eeqn
\[
\left[  \di, \dj \right] = 0.
\]
Then, we realize $U_i$ as
\[
U_1 = e^{i \sigma_1} e^{ 2\pi \theta \,  \partial_2},
~~U_2 = e^{i \sigma_2} e^{ -2\pi \theta \,  \partial_1}.
\]
For $\theta=0$  we have $U_i=\Ut_i=e^{i \sigma_i}$, all generators
commute allowing us to use either $U_i$'s or $\Ut_i$'s to generate
the algebra of functions on the classical torus.
It is then easy to check that 
\beqn
U_i^{-1}~\frac{1}{i}\dj~U_i = \frac{1}{i}\dj +\delta_{ij}.\label{UdU}
\eeqn
This and many other formulae in this paper can be proven using the 
Campbell-Baker-Hausdorff formula which can be written in closed form since
commutators like~(\ref{ss}) and~(\ref{ds}) are $c$-numbers. 
Equation~(\ref{UdU}) is 
very similar to the quotient condition~(\ref{quotient}) so one can write 
a solution as a sum of the partial derivative and a fluctuating 
part that commutes with 
the $U_i$'s. However this is just the definition of the
 covariant derivative
\beqarr
X_{i} &=& - 2\pi i R_j \delta_{i j} \dj + A_{i} (\Ut_i),  \label{con}  \\
X_{a} &=& X_{a} (\Ut_i), \nonumber  \\
\Theta~&=~& \Theta(\Ut_i), \nonumber
\eeqarr
where $A_i$, $X_a$ and each spinorial component of $\Theta$ 
are $n \times n$ hermitian matrices with operator
valued entries.
Note that since the partial derivatives already  satisfy the cocycle
condition, the gauge fields $A_i$ and the scalar fields $X_a$ must satisfy
a homogeneous quotient condition like the second relation 
in~(\ref{quotient}).
Hence $A_i$ and $X_a$ must depend only on  $\Ut_i$.  Hidden in this 
dependence is the fact 
that we are working on a trivial bundle over the quantum torus. 

If one
inserts~(\ref{con}) into the SQM action~(\ref{action}) the result is
a noncommutative SYM gauge field theory in $2+1$ dimensions, with
the space part given by the above quantum torus and a commutative time.
For the commutative case, matrix compactification on $T^{d}$ results in 
a SYM gauge theory in $d+1$ dimensions
on the dual torus. 
In the limit when the size of the original torus vanishes the dual
torus becomes ${\bf R}^{d}$, therefore we obtain the opposite of dimensional 
reduction.
If one starts from a Euclidean
10-dimensional SYM and dimensionally reduces in all directions
including the Euclidean time one obtains the IKKT~\cite{IKKT}
functional.
Matrix compactification of  one direction in the IKKT 
functional results in the finite temperature action
of the original theory~(\ref{action}).

\section{Twisted Quantum Bundles}
\label{TQB}
 
We can consider more general solutions of the quotient 
condition~(\ref{quotient}) which are connections on twisted bundles. 
They correspond to compactification of the DLCQ of M-theory
in the presence of transversely wrapped membranes. Again 
the solution is a sum of two terms, a constant curvature connection 
$D_i$ and a 
fluctuating part
\beqarr
X_{i} &=& - 2\pi i R_j \delta_{i j} D_j + A_{i} (Z_j), \label{twistcon}  \\
X_{a} &=& X_{a} (Z_i), \nonumber            \\
\Theta~ &=& ~\Theta (Z_i). \nonumber
\eeqarr
Here the $Z_i$'s are $n \times n$ matrices with operator entries and, 
just like the $\Ut_i$'s for the trivial bundle,  commute 
with the $U_i$'s, but
now are sections of the twisted bundle whose exact form will be 
discussed shortly. However, while for the trivial bundle
$A_i$, $X_a$ and the spinorial components of $\Theta$ are $n\times n$ matrix 
functions, in~(\ref{twistcon})  
$A_i$,$X_a$ and the components of $\Theta$ 
are one-dimensional functions but with matrix arguments.
Later, this will allow us to establish a relationship between a SYM on
a twisted $U(n)$ bundle and one on a $U(1)$ bundle. 

Following~\cite{PMH}, up to a gauge transformation the constant curvature
connection can be written as
\beqn
D_1 = \partial_1,~D_2=\partial_2 - i f \sigma_1, \label{Di}
\eeqn
where $f$ is the constant field strength
\[
[D_1,D_2 ] = - i f.
\]

Such a gauge field can only exist in a non-trivial bundle. One can 
introduce
transition functions $\Omega_i$ such that the sections of the 
fundamental bundle
satisfy the twisted boundary conditions
\beqarr
   \Phi(\sigma_1 + 2\pi,\sigma_2) &=& \Omega_1 (\sigma_1,\sigma_2) ~
   \Phi(\sigma_1,\sigma_2),                 \label{bcfund}   \\
   \Phi(\sigma_1,\sigma_2 + 2\pi) &=& \Omega_2 (\sigma_1,\sigma_2) ~
   \Phi(\sigma_1,\sigma_2). \nonumber
\eeqarr
Similarly the adjoint sections satisfy
\beqarr 
   \Psi(\sigma_1 + 2\pi,\sigma_2) = \Omega_1 (\sigma_1,\sigma_2) ~
   \Psi(\sigma_1,\sigma_2) 
~\Omega_1 (\sigma_1,\sigma_2)^{-1},   \label{bcadj}   \\
   \Psi(\sigma_1,\sigma_2 + 2\pi) = \Omega_2 (\sigma_1,\sigma_2) ~
   \Psi(\sigma_1,\sigma_2)   ~ \Omega_2 (\sigma_1,\sigma_2)^{-1}. \nonumber
\eeqarr
Consistency of the transition functions of the bundle requires that 
\beqn
\Omega_1(\sigma_1,\sigma_2 + 2\pi)~ \Omega_2(\sigma_1,\sigma_2) =
\Omega_2(\sigma_1+2\pi,\sigma_2)~\Omega_1(\sigma_1,\sigma_2). \label{cocycle}
\eeqn
This relation is known in the mathematical literature as the cocycle condition.
The covariant derivatives transform just as the adjoint
sections
\[ 
   D_i(\sigma_1 + 2\pi,\sigma_2) = \Omega_1 (\sigma_1,\sigma_2) ~
   D_i(\sigma_1,\sigma_2) ~\Omega_1 (\sigma_1,\sigma_2)^{-1},
\]
\[ 
   D_i(\sigma_1,\sigma_2 + 2\pi) = \Omega_2 (\sigma_1,\sigma_2) ~
   D_i(\sigma_1,\sigma_2)   ~ \Omega_2 (\sigma_1,\sigma_2)^{-1}.
\]
A particular solution for the transition functions compatible with the
constant curvature 
connection~(\ref{Di}) and satisfying the cocycle condition is given by 
\beqn
\Omega_1 = e^{i m \sigma_2 / n} U, ~ \Omega_2 =V, \label{trfunc}
\eeqn
where $U,~V$\/ are $n \times n$\/ unitary matrices satisfying
\[
UV=e^{-2\pi i m/n} VU
\]
and $m$ is an integer. For simplicity, here we will only consider
the case when $n$ and $m$ are relatively prime. For the general case
see~\cite{CDS,CK,BM}.
Using the representation  given in~\cite{PMH} one has
\[ 
U_{kl} = e^{2\pi i k m/n} \delta_{k,l},~~ V_{kl} = \delta_{k+1,l},
\]
where the subscripts are identified with period $n$.

We can express the above  matrices in terms of the standard 't~Hooft 
matrices~\cite{Hooft,Hooft2} denoted
here by $U'$ and $V'$ and satisfying
\[
U' V' = e^{-2\pi i/n} V' U',~~U'^{n}=V'^{n}=1.
\]
The relation is given by
\beqn
U= e^{2\pi i m/n}U'^m,~~ V=V'. \label{relation}
\eeqn
The phase in~(\ref{relation}) is due to the nonstandard definition
of $U$ used in~\cite{PMH}. This has certain advantages but
similar phases will appear when comparing the results 
of~\cite{PMH} with similar results where the standard 't~Hooft 
matrices were used.
We also introduce a unitary matrix $K$ which
changes the representation so that $V'$ is diagonal, and satisfies
\beqn
K U' K^{-1} = V'^{-1},~~K V' K^{-1} = U'.   \label{K}
\eeqn

Note that $n$ is quantized since we are considering a $U(n)$\/ gauge theory
and $m$ is quantized since the magnetic flux $f$ through $T_2$ is quantized
\[
2\pi f= \frac{m}{n-m\theta}.
\]
In M-theory $m$ is the transversal membrane wrapping number.

One can solve the boundary conditions~(\ref{bcfund}) for the fundamental
sections as in \cite{PMH} generalizing a previous result for $m=1$ in
the commutative case presented in \cite{GRT}. Using the ordered 
exponential explained below, the general solution has the form
\[
\Phi_k(\sigma_1,\sigma_2) = \sum_{s\in {\bf Z}} \sum_{j=1}^m
E\left(\frac{m}{n}\left(\frac{\sigma_2}{2\pi}+k+ns\right)+j,
i\sigma_1\right)
\widehat{\phi}_j \left(\frac{\sigma_2}{2\pi}+k+ns+\frac{nj}{m}\right).
\]
The ordered exponential~\cite{PMH} is defined 
for two variables whose commutator is a
$c$-number
\[
E(A,B)=\frac{1}{1-[A,B]}~\sum_{l=0}^{\infty} \frac{1}{l!} A^lB^l.
\] 
The normalization is such that
\[
E(-B,A)E(A,B) =1
\]
and it has the following desirable properties similar to the usual exponential
\beqarr
E(A+c,B)=E(A,B)e^{cB}, \label{prop} \\
E(A,B+c)=e^{cA}E(A,B). \nonumber
\eeqarr
The $\widehat{\phi}_j$ functions are defined on the whole real axis and are 
unrestricted except for the behavior at infinity. They should be considered 
as vectors in a Hilbert space on which all the elements of the algebra 
are represented.

Next we explain in some detail how to obtain this result. First we define
\[
\phi(\sigma_1,\sigma_2) \stackrel{\rm def}{=} \Phi_k 
(\sigma_1,\sigma_2-2\pi(k-1)).
\] 
The second boundary condition~(\ref{bcfund}) implies that
the definition of $\phi$ 
is consistent, i.e. $k$-independent. Using $V^n=1$ we also find that $\phi$ 
is a  periodic function
in $\sigma_2$
\[
\phi(\sigma_1,\sigma_2 +2\pi n) = \phi(\sigma_1,\sigma_2) .
\]
The other boundary condition gives
\[
\phi(\sigma_1+2\pi,\sigma_2) = e^{im(\sigma_2+2\pi)/n} 
\phi(\sigma_1,\sigma_2).
\]
It is convenient to separate out a factor to eliminate the above twist
\[
\phi(\sigma_1,\sigma_2)= f(\sigma_1,\sigma_2)
\check{\phi}(\sigma_1,\sigma_2)
\]
and to require a simpler periodicity condition for $\check{\phi}$
\[
\check{\phi}(\sigma_1+2\pi,\sigma_2) = \check{\phi}(\sigma_1,\sigma_2).
\] 
Then the function $f$ must satisfy
\[
 f(\sigma_1+2\pi,\sigma_2)=e^{i m(\sigma_2+2\pi)/n} f(\sigma_1,\sigma_2).
\]
This is satisfied exactly  for 
\[ f(\sigma_1,\sigma_2) = E \left( \frac{m}{n}\left(\frac{\sigma_2}{2\pi}
+1\right), i \sigma_1 \right),
\]
where in the right hand side we used the ordered exponential defined above.
Now we can Fourier transform $\check{\phi}$ in $\sigma_1$
\[
 \check{\phi}(\sigma_1,\sigma_2)=
\sum_{p \in {\bf Z}}~e^{ip\sigma_1}~\phi_p(\sigma_2)
\]
and using the property~(\ref{prop}) of the ordered exponential we obtain
\[
\phi(\sigma_1,\sigma_2)=\sum_{p \in {\bf Z}} E \left( \frac{m}{n}
\left(\frac{\sigma_2}{2\pi}
+1\right)+p, i \sigma_1 \right) \phi_p(\sigma_2).
\]
Let $p=ms+j$  with $j=1,\ldots ,m$ and $s$ is an integer. Then the solution 
can be
written as
\[
\phi(\sigma_1,\sigma_2)=\sum_{s \in {\bf Z}} \sum_{j=1}^{m} E \left( \frac{m}{n}
\left(\frac{\sigma_2}{2\pi}
+1\right)+ms+j, i \sigma_1 \right) \phi_{s,j}(\sigma_2),
\]
where $\phi_{s,j} \stackrel{\rm def}{=} \phi_{ms+j}$.
Periodicity in $\sigma_2$ then implies $\phi_{s-1,j}(\sigma_2 +2\pi n)=
\phi_{s,j}(\sigma_2)$ so that using this recursively we have
$\phi_{s,j}(\sigma_2)=\phi_{0,j}(\sigma_2+2\pi n s)$. Finally after defining
$\widetilde{\phi}_j(x) \stackrel{\rm def}{=} \phi_{0,j}(2\pi (x-1))$ 
we obtain
\[
\Phi_k(\sigma_1,\sigma_2)=\sum_{s \in {\bf Z}} \sum_{j=1}^{m}
 E \left( \frac{m}{n}
\left(\frac{\sigma_2}{2\pi}+k+ns\right)+j
, i \sigma_1 \right) \widetilde{\phi}_j\left(\frac{\sigma_2}{2\pi}+k+ns\right).
\]
This is the result mentioned above up to another redefinition
\[
\widetilde{\phi}_j(x)= \widehat{\phi}_j(x+\frac{n}{m}j).
\]

While the solutions for the sections of the fundamental bundle given
in~\cite{PMH} are suitable for showing the equivalence to the
projective modules of~~\cite{CDS} as we will 
discuss in Section~\ref{modules}, the appearance of the ordered exponential
is somewhat inconvenient. Using the special form of the transition 
functions we were able to rewrite the solution in an equivalent but  simpler 
form. The transition
functions in this gauge do not contain  $\sigma_1$\/ and it is convenient to 
order all $\sigma_1$\/ to the right in the solution. Using $V^n=1$\/ in 
the second condition~(\ref{bcfund}) one can express all $n$\/ components 
of $\Phi$  in terms of a single function with period $2\pi n$ in $\sigma_2$.
After Fourier transforming in $\sigma_2$ and imposing both boundary
conditions~(\ref{bcfund}) we obtain the general solution
\[
\Phi_k(\sigma_1,\sigma_2) = \sum_{p\in {\bf Z}}
~e^{2\pi i  (\sigma_2/2\pi+k) p/n}
~e^{2\pi i (\sigma_1/2\pi  -p/m)m/n}
~\widehat{\psi}_p(\sigma_1/2\pi -p/m),
\]
where only $m$ of the $\widehat{\psi}_p$ functions are independent, since
\[
\widehat{\psi}_{p+m}(x) = \widehat{\psi}_p(x).
\]

Using the same technique one can show that an arbitrary adjoint 
section  has the 
following  expansion
\beqn
\Psi(\sigma_1,\sigma_2)=\sum_{s,t \in {\bf Z}}~ c_{st} Z_1^s Z_2^{-t}. 
\label{adj}
\eeqn
Here $c_{s,t}$ are $c$-numbers and
\[
Z_1=e^{i\sigma_1/(n-m\theta)} V^b,~~Z_2=e^{i\sigma_2/n}U^{-b},
\]
where $b$ is an integer, such that we can find another integer $a$ satisfying
 $an-bm=1$. For $n$ and $m$ relatively prime one can always
find integer solutions to this equation. Again, we emphasize that the 
$Z_i$'s commute with the $U_i$'s.
They are generators
of the algebra of functions on a new quantum torus 
\[
Z_1 Z_2 = e^{2\pi i \theta'} Z_2 Z_1,
\]
where $\theta'$ is obtained by an $SL(2,{\bf Z})$ fractional 
transformation from $-\theta$
\[
\theta'=\frac{a(-\theta)+b}{m(-\theta)+n}.
\]

Now we outline how to obtain this result. 
Note first that
\[
\Psi(\sigma_1+2\pi n,\sigma_2) =\Omega_1^n ~\Psi(\sigma_1,\sigma_2)~
\Omega_1^{-n} = \Psi(\sigma_1+2\pi \theta m,\sigma_2).
\] 
In the last equality we used  the fact that $U^n =1$, and 
we also used the 
exponential  formula to shift $\sigma_1$. Using both
 boundary conditions we have
\[
\Psi(\sigma_1+2\pi(n-m\theta),\sigma_2)=\Psi(\sigma_1,\sigma_2),
\]
\[
\Psi(\sigma_1,\sigma_2+2\pi n)=\Psi(\sigma_1,\sigma_2).
\]
We can expand the section as
\[
\Psi(\sigma_1,\sigma_2)=\sum_{s,t\in {\bf Z}} e^{is\sigma_1/(n-m\theta)}
e^{-it\sigma_2/n} \Psi_{s,t},
\]
where $\Psi_{s,t}$ is a $n \times n$ matrix and can be expanded as
\beqn
\Psi_{s,t}=\sum_{i=i_0}^{n+i_0} \sum_{j=j_0}^{n+j_0} c_{s,t,i,j}V'^i U'^j.
\label{Psi}
\eeqn
Here $i_0,j_0$ are two arbitrary integers, allowing us to freely 
shift the summation 
limits assuming that $c_{s,t,i+n,j}=c_{s,t,i,j+n}=c_{s,t,i,j}$.
Then one can obtain further restrictions on the $c_{s,t,i,j}$ coefficients 
using the boundary conditions~(\ref{bcadj}). 
For example using the first equation~(\ref{bcadj}) and 
comparing like coefficients in the 
Fourier expansion we have
\[
c_{s,t,i,j} e^{2\pi i s/(n-m\theta)}= c_{s,t,i,j}  e^{-2\pi i mi/n}
e^{2\pi i s m\theta/[n(n-,\theta)]}.
\]
From this and the similar relation obtained by imposing  
the second equation~(\ref{bcadj}) we have that $c_{s,t,i,j}$ vanish unless
$(s+mi)/n=k$ and $(t+j)/n=s$ for $k$ and $s$ two integers.
These equations have multiple solutions. However, if $(i,j)$ and $(i',j')$ are 
two solutions then $i-i' \in n{\bf Z}$ and $j-j' \in n{\bf Z}$. This ensures 
that only
one term survives in the sum~(\ref{Psi}) over $i$ and $j$. Choosing
for later convenience
$i_0=sb$ and $j_0=mbt$ we have 
\[
\Psi(\sigma_1,\sigma_2)=\sum_{s,t\in {\bf Z}}~
 e^{is\sigma_1/(n-m\theta)}
e^{-it\sigma_2/n}  \sum_{i=sb}^{n+sb}~
 \sum_{j=mbt}^{n+mbt} c_{s,t,i,j}~V'^{i} U'^{j} .
\]
Since $n$ and $m$ are
relatively prime let $a,b \in {\bf Z}$ such that $an-bm=1$. Then
\[
k=as,~l=at,~i=bs,~j=mbt
\]
is an integer solution inside the $i$ and $j$ summation range. 
Dropping the $i,j$ indices since they are determined 
by $s$ and $t$ we have
\[
\Psi(\sigma_1,\sigma_2) = \sum_{s,t \in {\bf Z}} 
c_{s,t} \left(   e^{i\sigma_1/(n-m\theta)} V'^b    \right)^{s}
\left(  e^{i\sigma_2/n} U'^{-mb}  \right)^{-t},
\]
which is just~(\ref{adj}) after an additional phase redefinition of 
$c_{s,t}$ to accommodate the phase difference between $U$ and $U'^m$.

\section{Projective Modules and Morita Equivalence}
\label{modules}

A classic mathematical result of Gel'fand states that compact 
topological spaces are in one to one correspondence with 
commutative $C^{*}$-algebras. In one direction, to a topological space $X$
we associate the algebra of continuous functions $C(X)$.
Conversely and rather nontrivially, 
the spectrum of a commutative $C^{*}$-algebra is equivalent 
to a compact topological space. This important result 
allows for a dual description of topological spaces and brings 
powerful algebraic methods into the realm of topology. On the other hand,
if we drop the commutativity requirement, a $C^{*}$-algebra 
${\cal A}$ describe
what is called by correspondence a quantum space. To illustrate, consider
the algebra of the quantum torus ${\cal A}(T^2_{\theta})$ generated 
by the $U_i$'s. An arbitrary
element $a$ has the form
\beqn
a = \sum_{k,l\in {\bf Z}} a_{k,l} U_{1}^k U_{2}^{l}, \label{expand}
\eeqn
where some restrictions (which we do not discuss here) are imposed on
the $c$-number coefficients $a_{k,l}$. For $\theta=0$, 
formula~(\ref{expand}) reduces to the Fourier expansion of
functions on $T^2$. Thus we can read the compact space from the
commutative algebra. 

Using the same strategy one can describe other spaces of classical geometry
in commutative algebraic terms and then remove the commutativity
requirement. A quantum vector bundle is a 
projective ${\cal A}$-module ${\cal E}$. First consider the classical 
commutative picture. The set ${\cal E}$ of  global sections
of a vector bundle over a base space $X$ has the structure of a projective
module over the algebra $C(X)$. Having a module 
essentially means that we can add sections
and can multiply them by functions. Not all modules over a commutative
algebra are vector bundles. For example the set of sections on a
space consisting of a collection of fibers of different 
dimensions over a base space also form a module. However,
projective modules over the
algebra of functions on a topological space 
are in one to one correspondence with vector bundles over that space.
By definition a projective module is a direct summand in a free module.
A free module ${\cal E}_0$ over an algebra ${\cal A}$ is a module 
isomorphic to a direct sum of a finite number of copies of the algebra
\[
{\cal E}_0={\cal A}\oplus \ldots \oplus{\cal A}.
\] 
Trivial bundles
correspond to free modules since the description 
of their sections in terms of components
is global, and each component is an element of $C(X)$. 
For every vector bundle  we can find
another one such that their direct sum is a trivial bundle.
In dual language this implies that the module of sections ${\cal E}$ is 
projective
\[
{\cal E}_0 = {\cal E} \oplus {\cal E}'.
\]
Again it is nontrivial to show the converse, that every projective module 
is isomorphic to the set of sections of some vector bundle.
Finally projective modules over noncommutative algebras are 
the quantum version of vector bundles.

In the noncommutative case we distinguish between left and right 
projective modules. Multiplying fundamental sections from the right
with elements of ${\cal A}(T^2_{- \theta})$ preserves the 
boundary conditions~(\ref{bcfund}) while multiplication on the left
gives something that no longer is a global section. Thus the set of
sections of the fundamental bundle form a right projective module
over the ${\cal A}(T^2_{- \theta})$ algebra 
which we denote ${\cal F}_{n,m}^{\theta}$.
This is no longer true for the 
adjoint sections since in~(\ref{bcadj}) the transition functions 
multiply from both the left and right. However one can check that
the fundamental and the adjoint 
are both left and right
projective modules over the ${\cal A}(T^2_{\theta})$ algebra.
This is because the exponents of the $U_i$'s satisfy 
\beqn
[i\sigma_1+2\pi \theta \partial_2,\sigma_i]=0,
~~[i\sigma_2-2\pi \theta \partial_1,\sigma_i]=0; \label{Uscomute}
\eeqn
thus the $U_i$'s can be commuted over the transition 
functions in~(\ref{bcfund}) and ~(\ref{bcadj}). 
Additionally, the fact that ${\cal F}_{n,m}^{\theta}$
is both a 
left ${\cal A}(T^{2}_{\theta})$-module and a 
right ${\cal A}(T^{2}_{-\theta})$-module can be understood as follows.
Since $[U_i,\sigma_j]=0$ we have
\[
U_i \Phi(\sigma_1,\sigma_2)=\Phi(\sigma_1,\sigma_2) \Ut_i,
\]
where we dropped the derivatives when there was nothing to their right.
Thus multiplying on the left with $a$
is equivalent
to multiplying on the right with $\widetilde{a}$  
\beqn
a \Phi = \Phi \widetilde{a},    \label{swich}
\eeqn
where $\widetilde{a} = \sum_{k,l\in {\bf Z}} a_{k,l} \widetilde{U}_{2}^{l}  
\widetilde{U}_{1}^k$ is the same function as $a$ but with $\Ut_i$'s
as arguments and with all the factors written in reversed order.

As mentioned in \cite{PMH} the construction in Section~\ref{SYM}
is equivalent to the projective
modules discussed in \cite{CDS}. 
By solving the boundary conditions we went from a local 
to a global description.
Here we present
explicit formulae for this equivalence. First one has to express the 
left actions  on the fundamental sections as 
actions on the Hilbert space \cite{PMH}. For example the action 
of the $Z_i$ generators is given by
\[
(Z_1 \widehat{\phi})_j(x)=\widehat{\phi}_{j-a}(x-\frac{1}{m}),~~
(Z_2 \widehat{\phi})_j(x)= e^{-2\pi i j/m}e^{2\pi i x/(n-m\theta)}
\widehat{\phi}_{j}(x).
\]
This can be written as
\[
Z_1=W_1^a V_1,~~Z_2=W_2 V_2,
\]
where $V_i$ and $W_i$ are operators acting on the Hilbert space as
\[
(V_1 \widehat{\phi})_j(x)=\widehat{\phi}_j(x-\frac{1}{m}),~~   
(V_2 \widehat{\phi})_j(x)=e^{2\pi i x/(n-m\theta)} \widehat{\phi}_j(x), 
\]
\[
(W_1 \widehat{\phi})_j(x)= \widehat{\phi}_{j-1}(x),~~
(W_2 \widehat{\phi})_j(x)=e^{-2\pi i j /m}\widehat{\phi}_j(x). 
\]
These operators satisfy the following relations
\[
V_1 V_2 = e^{-2\pi i/[m(n-m\theta)]} V_2 V_1,~~
W_1 W_2 = e^{2\pi i/m} W_2 W_1,~~[V_i,W_j]=0
\]
and can be used to express other operators acting in the Hilbert space.
For example we have 
$U_1=W_1 V_1^{n-m\theta}$ and $U_2=W_2^{n}V_2^{n-m\theta}$.

We can now present the correspondence between \cite{CDS} and \cite{PMH}.
The two integers $p$ and $q$ and the angular variable
 $\theta_{\rm CDS}$ labeling the projective module 
${\cal H}_{p,q}^{\theta_{\rm CDS}}$ of~\cite{CDS},
and $\theta'_{\rm CDS}$ 
can be expressed 
in terms of the quantities used in this paper or in~\cite{PMH}
\[
p=n,~~q=-m,~~\theta_{\rm CDS}=-\theta,~~\theta'_{\rm CDS}=\theta'.
\]
Then $ {\cal F}_{n,m}^{\theta}  \cong {\cal H}_{n,-m}^{-\theta}$.
The Hilbert space representation of~\cite{CDS} written in terms of
the function $f(s,k)$ with $s \in {\bf R}$ and $k \in {\bf Z}_q$ is linearly
related to the $\widehat{\phi}_k(x)$ representation
\[
\widehat{\phi}_k(x)=
\sum_{l=1}^{m} 
{\cal K}_{kl}~({\cal S}_{(-\frac{m}{n-m\theta})} f)(x,l).
\]
Here ${\cal K}$ is an $m\times m$ representation changing matrix 
defined as in~(\ref{K}) but for $m$-dimensional 't~Hooft matrices, and
${\cal S}_{\lambda}$ is the rescaling operator 
$({\cal S}_{\lambda}f)(x,k)=f(\lambda x,k)$ which can be expressed using the 
ordered exponential
\[
{\cal S}_{\lambda}= \lambda E((\lambda-1)x,\partial_x).
\]
Also, using lower case 
to distinguish them from our current notation which 
follows \cite{PMH}, the operators in \cite{CDS} represented in
the $\hat{\phi}_k(x)$ basis are given by
\[
v_0= V_2^{n-m\theta},~~v_1=V_1^{n-m\theta},
~~w_0=e^{2\pi i n/m} W_2^{n},~~w_1=e^{2\pi i /m} W_1
\]
\[
z_0=e^{2\pi i /m} Z_2,~~z_1=e^{-2\pi i a/m} Z_1^{-1},
~~u_0=e^{2\pi i n/m} U_2,~~u_1=e^{2\pi i /m} U_1.
\]

Next we introduce the Morita equivalence of two 
algebras~\cite{R1,R2,R3,R4,AS1,AS2}, which 
can be used
to describe a subgroup of the T-duality group of the 
M-theory compactification in the language of noncommutative SYM gauge theory. 

Two $C^{*}$-algebras ${\cal A}$ and ${\cal A}'$ are 
Morita equivalent 
if there exists a right ${\cal A}$-module ${\cal E}$ such that the algebra
$End_{\cal A}{\cal E}$ is isomorphic to ${\cal A}'$. Here 
$End_{\cal A}{\cal E}$ denotes the set of endomorphisms of 
the ${\cal A}$-module ${\cal E}$. It consists of linear 
maps $T$ on ${\cal E}$ where linearity is not only with respect to
$c$-numbers but also with respect to right multiplication by 
elements of ${\cal A}$
\[
T(\Phi f) = T(\Phi) f,~~\Phi \in {\cal E},~~f \in {\cal A}.
\]

An example of Morita equivalent algebras is
 ${\cal A}(T^{2}_{-\theta})$ 
and ${\cal A}(T^{2}_{\theta'})$. 
As discussed above,
the projective module associated to the quantum fundamental bundle 
${\cal F}_{n,m}^{\theta}$ is
a right ${\cal A}(T^{2}_{-\theta})$-module. 
One can prove that
$End_{{\cal A}(T^{2}_{-\theta})} {\cal F}_{n,m}^{\theta}$ is isomorphic to 
${\cal A}(T^{2}_{\theta'})$.
Here we just show that the two algebras have the same generators.
Using~(\ref{swich})  we have $T(\Phi \widetilde{a})=T(a \Phi)$ and 
$T(\Phi)\widetilde{a}=a T(\phi)$
and since $T$ is an endomorphism we obtain
$T(a \Phi)=a T(\phi)$, which can also be
written as $[T,a]=0$. But the $Z_i$'s were found exactly by requiring 
that they 
commute with $U_i$'s so $T \in {\cal A}(T^2_{\theta'})$.  

The physical interpretation of Morita equivalence is that 
a $U(n)$ SYM gauge theory on the twisted bundle with magnetic
flux $m$ is equivalent to a $U(1)$ gauge theory on a dual
quantum torus $T^2_{\theta'}$. 
This can be seen as a consequence of the discussion following 
equation~(\ref{twistcon}).
The gauge field components $A_i$, the
scalar fields $X_a$, and the components of $\Theta$ are not matrix
valued, rather they are one-dimensional. The final 
result is a matrix because the $Z_i$'s are matrices. On the other hand,
we can ignore the internal structure of the $Z_i$'s and just regard them as the
generators of ${\cal A}(T^{2}_{\theta'})$, thus
allowing us to reinterpret the original theory as a
noncommutative $U(1)$ gauge theory
on the quantum torus ${\cal A}(T^{2}_{\theta'})$.

Generally, two theories with parameters
$(n_1,m_1,\theta_1)$ and $(n_2,m_2,\theta_2)$ and appropriately chosen
compactification radii are equivalent if they are on the same orbit of the 
$SL(2,{\bf Z})$ duality group
\[
-\theta_2=\frac{A(-\theta_1)+B}{C(-\theta_1)+D},
\]
\[
\left( \begin{array}{c}  n_2 \\-m_2 \end{array} \right)
             =
\left( \begin{array}{cc} A & B \\ C & D \end{array} \right) 
\left( \begin{array}{c} n_1 \\-m_1 \end{array} \right) .
\]
Since $\theta$ is a continuous variable, we can interpolate
continuously, through noncommutative SYM theories, between two
commutative SYM theories with gauge groups of different rank and appropriate
magnetic fluxes. This $SL(2,{\bf Z})$ duality subgroup has a nice geometric 
interpretation in the T-dual picture of~\cite{DH} where it corresponds to a 
change of basis of the dual torus lattice~\cite{CK,BM}.

\section{Gauge Transformations}
\label{gauge-trans}

In this section we consider a gauge equivalent formulation of the previous
results closely following the treatment of Taylor in~\cite{WT} 
of the corresponding commutative case. In that paper a gauge transformation 
was considered so as to change the standard 't~Hooft transition function
into trivial transition function in the $X_2$ direction. When the 
transition functions are trivial T-duality has the standard form,
i.e. the gauge field translates directly into the position of a
D-string on the dual torus. Here we show that a similar gauge 
transformation can be performed in the noncommutative case. See 
also~\cite{BM}
for further discussions of this including a relation to the three-string
vertex of Douglas and Hull introduced in~\cite{DH}.

First let us consider a general gauge transformation $g(\sigma_1,\sigma_2)$.
Just as in the classical case  the covariant
derivatives transform as $D'_i= g^{-1} D_i g \label{D'}$ 
resulting in the following transformation
for the gauge fields
\beqn
A'_i = g^{-1} A_i g + i g^{-1} \partial_i g. \label{A'}
\eeqn
As a result the new transition functions are given by 
\beqarr
\Omega'_1(\sigma_1,\sigma_2) = g^{-1}(\sigma_1+2\pi,\sigma_2)
\Omega_1(\sigma_1,\sigma_2) g(\sigma_1,\sigma_2), \label{O'}  \\
\Omega'_2(\sigma_1,\sigma_2) = g^{-1}(\sigma_1,\sigma_2+2\pi)
\Omega_2(\sigma_1,\sigma_2) g(\sigma_1,\sigma_2). \nonumber
\eeqarr
Again all this is just as in the classical case except that one has to take 
into account the noncommutativity of the $\sigma_i$'s.

It will be useful to consider first the $\theta=0$ commutative case. 
Then we know
both the original gauge fields~(\ref{Di}) and the transformed ones 
\[
A'_1=0,~~A'_2=\frac{m}{n}\frac{\sigma_1}{2\pi} + Q,
\]
where $Q=\frac{1}{n} {\rm diag}(0,1,\ldots,n-1)$, and we use primes for all 
variables in the new gauge. We can write a 
differential equation for the gauge transformation
\[
\partial_1 g=0,~~i \partial_2 g = g A'_2 - A_2g = g Q
\] 
which can be integrated to give 
\beqn
g=K e^{-i\sigma_2 Q} \label{gauge},
\eeqn
where the integration constant $K$ is the $n\times n$ matrix~(\ref{K}).
It was fixed by requiring 
a trivial $\Omega'_2$ as given by~(\ref{O'}). 
Using the gauge transformation~(\ref{gauge}) we can now calculate both 
transition functions
\beqn
\Omega'_1 = e^{2\pi i m/n} e^{i\sigma_2 T_m} V^m,~~\Omega'_2 =1 \label{trfunc'}
\eeqn
where $T_k={\rm diag}(0,\ldots,0,1,\ldots,1),~k=1,\ldots,n$ with 
the first $n-k$ 
entries vanishing and the last $k$ equal to unity. 

Next we discuss the noncommutative case. The first thing to notice is that
the original quantum transition functions~(\ref{trfunc}) are $\theta$
independent and only contain the $\sigma_2$ variable. Similarly the 
classical gauge transformation~(\ref{gauge}) only depends on $\sigma_2$ so
that the classical computation of the new transition functions is also 
valid in the quantum case.
Using~(\ref{A'}) we can compute the new gauge fields
\[
A'_1=0,~~
A'_2=\frac{m}{n-m\theta}\frac{\sigma_1}{2\pi}+\frac{n}{n-m\theta}~Q.
\]

Since~(\ref{Uscomute}) implies $[U_i,g]=0$ we see that
the gauge transformation is compatible with the 
quotient conditions~(\ref{quotient}).
We can use the gauge transformation to obtain the generators
of the sections of the adjoint bundle
\[
Z'_1 = e^{i\sigma_1/(n-m\theta)} e^{-2\pi i n \theta' Q},
~~Z'_2= e^{2\pi i/n}V e^{i\sigma_2(1-T_{n-1})}.
\]
We can also write explicit formulae for the fundamental sections in the
new gauge
\[
\Phi'_k(\sigma_1,\sigma_2) =
\sum_{r \in {\bf Z}} e^{i\sigma_2 r} \chi_{k-nr}\left( \frac{\sigma_1}{2\pi}+
\frac{k-nr}{m} \right).
\]
The $\chi_s$ functions are defined over the real axis and must satisfy
\[
\chi_{s+m}(x)=e^{-2\pi i m/n} \chi_{s}(x),
\]
so that only $m$ of them are independent.
Again, we note that since the transition functions only contain 
$\sigma_2$ and all were ordered to the left of $\sigma_1$ in the 
solution for the sections of the fundamental bundle, they have the same form
in the noncommutative and in the classical case.

\section*{Acknowledgments}
It is a pleasure to dedicate this paper to Dick Arnowitt.
We would like to thank Korkut
Bardakci and Paolo Aschieri for useful discussions
and valuable comments.
This work was supported in part by 
the Director, Office of Energy Research, Office of High Energy and Nuclear
Physics, Division of High Energy Physics of the U.S. Department of Energy
under Contract DE-AC03-76SF00098 and in part by the National Science 
Foundation under grant PHY-95-14797.


\end{document}